\documentclass[10pt,conference]{IEEEtran}
\usepackage{amsmath,amssymb,url,graphicx,epsfig}

\title{Channel Capacity Limits of Cognitive Radio in Asymmetric Fading Environments}

\begin{document}
\newcounter{MYtempeqncnt}
\pagestyle{empty}
\thispagestyle{empty}
%
%
%
\author{\authorblockA{Himal A. Suraweera~\IEEEauthorrefmark{1}, Jason Gao~\IEEEauthorrefmark{1}, Peter J. Smith~\IEEEauthorrefmark{2}, Mansoor Shafi~\IEEEauthorrefmark{3} and Michael Faulkner~\IEEEauthorrefmark{1}\\
\IEEEauthorrefmark{1}~School of Electrical Engineering, Victoria University, Melbourne, Australia\\
}
\authorblockA{\IEEEauthorrefmark{2}Department of Electrical and Computer Engineering, University of Canterbury, Christchurch, New Zealand}
\authorblockA{\IEEEauthorrefmark{3}~Telecom New Zealand, PO Box 293, Wellington, New Zealand\\ Email: himal.suraweera@vu.edu.au,~p.smith@elec.canterbury.ac.nz,~mansoor.shafi@telecom.co.nz}}
\maketitle
\thispagestyle{empty}
\pagestyle{empty}
\begin{abstract}
Cognitive radio technology is an innovative radio design concept which aims to increase spectrum utilization by exploiting unused spectrum in dynamically changing environments. By extending previous results, we investigate the capacity gains achievable with this dynamic spectrum approach in asymmetric fading channels. More specifically, we allow the secondary-to-primary and secondary-to-secondary user channels to undergo Rayleigh or Rician fading, with arbitrary link power. In order to compute the capacity, we derive the distributions of ratios of Rayleigh and Rician variables. Compared to the symmetric fading scenario, our results indicate several interesting features of the capacity behaviour under both average and peak received power constraints. Finally, the impact of multiple primary users on the capacity under asymmetric fading has also been studied.
\end{abstract}
\section{Introduction}
Conservative spectrum policies employed by regulatory authorities have resulted in spectrum underutilization of the overall available spectrum for wireless communications. Measurements performed by agencies such as the Federal Communications Commission \cite{fcc} in the United States and Ofcom \cite{ofcom} in the United Kingdom have revealed that at any given time, large portions of spectrum are sparsely occupied. Findings of such campaigns on spectrum usage have challenged the traditional spectrum management approaches.

The concept of cognitive radio (CR) \cite{mitola} refers to a smart radio which can sense the external electromagnetic environment and adapt its transmission parameters according to the current state of the environment. CRs can access parts of the spectrum for their information transmission, provided that they cause minimal interference to the primary users in that band~\cite{jondral,akyildiz}. Therefore, spectrum sharing among the primary licensee and the secondary CR must be carried out in a controlled fashion. In the technical literature, the \emph{interference temperature} introduced by Kolodzy \cite{kolodzy,haykin} indicates the interference level at the primary licensee's receiver. From the licensees' point of view, the secondary access can be controlled in two ways. The total interference power can be required to remain below a certain threshold (an interference temperature constraint) or the signal-to-noise-and-interference (SINR) can be constrained. 

The capacity of wireless systems has been extensively studied under fixed spectrum access. For CR, this work is less mature and many information/communication theoretic problems and implementation issues \cite{cabric} remain to be solved. However, several interesting results on the capacity, outage probability and throughput of CR systems have recently emerged. See for example,~\cite{ghasemi_b,jafar,gastpar_1}. In \cite{gastpar_1}, Gastpar derived the capacity of different non-fading additive-white-Gaussian-noise (AWGN) channels with the average received-power at a primary receiver being constrained. In \cite{ghasemi_b}, Ghasemi and Sousa showed that with the same limit on the received-power level, channel capacity for a range of fading models (e.g., Rayleigh, Nakagami-$m$ and log-normal fading) exceeds that of the non-fading AWGN channel. In some scenarios, primary user spectral activity in the vicinity of the cognitive transmitter may differ from that in the vicinity of the cognitive receiver. Considering this, in \cite{jafar}, the capacity of opportunistic spectrum acquisition in the presence of distributed spectral activity has been investigated.

We extend the work in \cite{ghasemi_b} which assumed that fading conditions for the interference path (CR transmitter-primary receiver) and the desired path (CR transmitter-CR receiver) are the same. In practice, these two links could experience different fading conditions (types) and different link powers (due to path length or shadowing). This is referred as \emph{asymmetric} fading in this paper.

In this paper we consider Rayleigh and Rician fading. Hence, we are able to quantify the effects of propagation paths consisting of both line-of-sight (LoS) and scattered components. Building on the work
in \cite{ghasemi_b}, this paper makes the following main contributions:
\begin{enumerate}
\item
The secondary capacity under average-received power and peak-power constraints is studied for asymmetric conditions including different fading types (Rayleigh and Rician) and different link powers. Here we show that under low interference to the primary receiver, the secondary capacity is sensitive to the fading type on the desired and interference paths.
\item
The impact of multiple primary licensee receivers in Rayleigh and Rician fading is studied for
peak power constraints.
\item
Closed-form expressions for the cumulative distribution function
(CDF) and the probability density function (PDF) of a random
variable (RV), $g_1/g_0$ is derived, for the cases where
$(\sqrt{g_1},\sqrt{g_0})$ experience (Rayleigh, Rician) and (Rician,
Rayleigh) fading. This is needed to derive the above mentioned
capacities.

\end{enumerate}
This paper is organized as follows. The system and channel model is
described in Section~II. In Section~III we derive the exact PDFs for
the ratio of Rayleigh and Rician RVs. In Section~IV, these results
are used to study the capacity gains under average/peak
received-power constraints respectively. Extensions of these results
to multiple primary users are presented in Section~V. Finally, some
conclusions are drawn in Section~VI. Throughout the paper, the
reference to average/peak received-power refers to as the
average/peak interference power at the primary receiver. The CR link
is also referred to as a secondary link.
\begin{figure}[t]
\centering
\includegraphics[width=0.85\columnwidth]{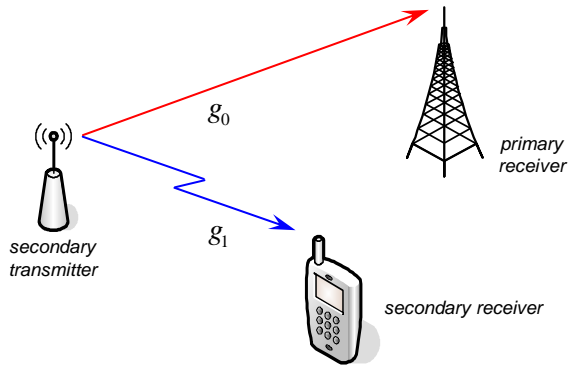}
\caption{Shared spectrum usage between primary and secondary users.}
\label{fig_1}
\vspace{-3mm}
\end{figure}
\section{System and Channel Model}
In this section, the system and channel model considered in the paper are briefly outlined (cf. Fig. 1). The system model is borrowed from \cite{ghasemi_b}, however we have considered asymmetric fading scenarios. A point-to-point flat fading channel with perfect channel side information available to both the receiver and the
transmitter is assumed. Let $g_0$ and $g_1$ denote the instantaneous channel gains from the secondary transmitter to the primary and secondary receivers respectively. Furthermore, we denote the respective PDFs by $p_{g_0}(g_0)$ and $p_{g_1}(g_1)$. For a unit power channel gain, the Rayleigh PDF is given by
\begin{equation}
p_{\sqrt{g}}(x)=2xe^{-x^2}
\end{equation}
for $x \ge 0$. For a Rician distribution the PDF is given by
\begin{align}
p_{\sqrt{g}}(x)=2x(1+K)e^{-K-(1+K)x^2}I_{0}\left(2x\sqrt{K+K^2}\right)
\end{align}
for $x \ge 0$, where $K$ is the Rician $K$-factor defined as the ratio of signal power in dominant component over the scattered power and $I_0(\cdot)$ is the zeroth-order modified Bessel function of the first kind. For $K=0$, Rayleigh fading is experienced and $K=\infty$ gives the AWGN (no fading) situation. Values of the $K$-factor in indoor/outdoor land mobile applications normally range from $0-12$ dB \cite{parsons}. In a practical environment the CR transmitter to CR receiver link may not be of the same length as the CR interference path to the primary receiver. When the link powers, $E\{g_0\}$ and $E\{g_1\}$, differ, it can be shown that capacity only depends on the power ratio. Hence, we define the relative power parameter, $c$, by $c=E\{g_1\}/E\{g_0\}$. Note that $E\{\cdot\}$ denotes the expectation operator.

Two important items of notation should be stressed at this point. Since the main results of the paper depend on the ratio $g_1/g_0$, we use the shorthand notation Rayleigh/Rician to indicate that $\sqrt{g_1}$ is Rayleigh and $\sqrt{g_0}$ is Rician. Similarly, Rician/Rayleigh indicates that $\sqrt{g_1}$ is Rician and $\sqrt{g_0}$ is Rayleigh. The second issue is that the secondary transmitter must constrain its power so that the interference at the primary is acceptable. Hence the power constraints in this scenario are really interference constraints. This is different to many other problems where the constraints are for transmit power.
\section{CDF and PDF of $g_1/g_0$}
Here, we anticipate the results of Section~IV, where it is shown that capacity depends on the ratio, $g_1/g_0$. Hence, in this section the CDF and the PDF for a Rayleigh/Rician RV and a Rician/Rayleigh RV are derived. 

Consider the distribution of a Rayleigh/Rician RV, $X=g_1/g_0$. Mathematically, $P(X<x)$, i.e., the CDF of $X$, is given by
\begin{equation}
\label{ae1}
P(X<x)=\int^{\infty}_{0}P\left(\frac{g_1}{g_0}<x|g_0\right)p_{g_0}(g_0)dg_0
\end{equation}
Equation \eqref{ae1} can be simplified as
\begin{align}
\label{re1}
P(X<x)& =(K+1)e^{-K} \int^{\infty}_{0}(1-e^{-xg_0})\\\nonumber
& \cdot e^{-(K+1)g_0}I_{0}(2\sqrt{K(K+1)g_0})dg_0
\end{align}
The integral in \eqref{re1} can be solved using \cite[eq. 2.15.5.4]{prud} and we obtain the CDF of $X$ as
\begin{align}
\label{ty1}
F_{X}(x) = 1-\frac{K+1}{x+K+1}e^{-K+\frac{K^2+K}{x+K+1}}
\end{align}
for $x \ge 0$. The PDF of $X$ can be found by taking the derivative of \eqref{ty1} with respect to $x$, yielding
\begin{equation}
\label{rayrician}
p_{X}(x)=(K+1)\frac{x+(K+1)^2}{(x+K+1)^3}e^{-K+\frac{K^2+K}{x+K+1}}
\end{equation}
for $x \ge 0$. As expected, for $K=0$ the PDF $p_{X}(x)$ is given by $p_{X}(x)=1/(x+1)^2$ \cite[eq. 11]{ghasemi_b}.

Now consider the distribution of $Y=g_1/g_0$ when $\sqrt{g_1}$ is Rician and $\sqrt{g_0}$ is Rayleigh. Using the same approach, $P(Y<y)$ is given by
\begin{align}
\label{iy1}
P(Y<y)=1-\int^{\infty}_{0}Q_1\left(\sqrt{2K},\sqrt{2(1+K)yg_0}\right)e^{-g_0}dg_0
\end{align}
where $Q_{1}(a,b)=\int^{\infty}_{b}xe^{-\frac{a^2+b^2}{2}}I_{0}(a x)dx$ is the first-order Marcum $Q$-function which satisfies the following identity \cite[eq. 5]{nuttall}
\begin{align}
\label{h1}
Q_1(a,b)+Q_1(b,a)=1+e^{-\frac{a^2+b^2}{2}}I_{0}(ab)
\end{align}
Using \eqref{h1}, we can express \eqref{iy1} as shown in \eqref{inr4}.
\begin{figure*}[!t]
\normalsize
\setcounter{MYtempeqncnt}{\value{equation}}
\setcounter{equation}{8}
\begin{align}
\label{inr4}
P(Y<y)=-\int^{\infty}_{0}e^{-(y+Ky+1)g_0}I_{0}(2\sqrt{(Ky+K^2y)g_0})dg_0+\int^{\infty}_{0}Q_{1}(\sqrt{2(y+Ky)g_0},\sqrt{2K})e^{-g_0}dg_0
\end{align}
\setcounter{equation}{\value{MYtempeqncnt}}
\hrulefill
\end{figure*}
\addtocounter{equation}{1}
The first integral in \eqref{inr4} can be evaluated in closed-form using the result of \cite[eq. 2.15.5.4]{prud} and is
\begin{align}
I_1=\frac {e^{-K+\frac{Ky+K^2y}{y+Ky+1}}}{{y+Ky+1}}
\end{align}
The second integral in \eqref{inr4} can be evaluated using the result of \cite[eq. 25]{nuttall_2} as $I_2=e^{-\frac{K}{y+Ky+1}}$. Hence, $P(Y<y)$ is given by
\begin{align}
\label{yr1}
F_{Y}(y)=e^{-\frac{K}{y+Ky+1}}-\frac {e^{-K+\frac{Ky+K^2y}{y+Ky+1}}}{{y+Ky+1}}
\end{align}
for $y \ge 0$. After taking the derivative of \eqref{yr1} with respect to $y$, we obtain the PDF of $Y$ as
\begin{align}
\label{ricianray}
p_{Y}(y)& = \frac{K(1+K)}{(y+Ky+1)^2}e^{-\frac{K}{y+Ky+1}}\\\nonumber
& +\frac{(1+K)^2(1-K+y)}{(y+Ky+1)^3}e^{-K+\frac{Ky+K^2y}{y+Ky+1}}
\end{align}
for $y \ge 0$. For Rayleigh/Rayleigh fading, the PDF $p_{Y}(y)$ simplifies to $p_{Y}(y)=1/(y+1)^2$ \cite[eq. 11]{ghasemi_b}. To confirm the derivations, the PDFs of \eqref{rayrician} and \eqref{ricianray} were validated by Monte Carlo simulations, and a perfect match was obtained.
\section{Capacity Gains of Spectrum Sharing}
\subsection{Capacity Under an Average Received-Power Constraint}
In this section, we investigate the capacity gains achievable by the secondary user under an average received-power constraint. In \cite{ghasemi_b}, the channel capacity was expressed as
\begin{align}
\label{first}
C=\int\int B\log_2\left(1+\frac{g_1 P(g_0,g_1)}{N_0 B}\right)p_{g_0}(g_0)p_{g_1}(g_1)dg_0dg_1
\end{align}
such that
\begin{align}
\int_{0}^{\infty}\int_{0}^{\infty}g_0P(g_0,g_1)p_{g_0}(g_0)p_{g_1}(g_1)dg_0dg_1 \leq Q
\end{align}
where $Q$ is the maximum average interference power tolerated by the primary receiver~\footnotemark[2]\footnotetext[2]{The quality of transmission at the primary receiver can also be measured using the SINR. This requires a knowledge of the primary transmitter to primary receiver channel.}, $B$ is the available bandwidth, $N_0$ is the noise power at the secondary receiver and $P(g_0,g_1)$ denotes the optimal power allocation. Using the Lagrangian technique, \cite{ghasemi_b} has found $P(g_0,g_1)$ to be
\begin{equation}
P(g_0,g_1)=\left(\frac{1}{\lambda_0g_0}-\frac{N_0 B}{g_1}\right)^+
\end{equation}
where $(\cdot)^{+}$ denotes $\max\{\cdot,0\}$. Note that $\lambda_0$ is determined such that the average receive power is equal to $Q$. That is mathematically
\begin{equation}
\label{u3}
\int_{g_0}\int_{g_1}\left(\frac{1}{\lambda_0}-N_{0}B\frac{g_0}{g_1}\right)^{+}p_{g_1}(g_1)p_{g_0}(g_0)dg_1dg_0=Q
\end{equation}
Hence, the channel capacity can be calculated from
\begin{equation}
\label{u1}
C=B\int^{\infty}_{\frac{1}{\gamma_0}}\log_2(\gamma_0g_{10})p_{\frac{g_1}{g_0}}(g_{10})dg_{10}
\end{equation}
where $\gamma_0=1/(\lambda_0N_{0} B)$ and $p_{\frac{g_1}{g_0}}(\cdot)$ denotes the PDF of $g_1/g_0$. To the best of the authors' knowledge, there are no closed-form solutions for the integral in \eqref{u3} for the two fading scenarios considered in this paper. Rewriting    \eqref{u3} gives
\begin{align}
\label{fd1}
\int^{\gamma_0}_{0}(\gamma_0-x)p(x)dx=\frac{Q}{N_0B}=\alpha
\end{align}
where $p(x)$ in \eqref{fd1} denotes the PDF of $g_0/g_1$. Therefore, $\alpha$ is the allowable interference-to-noise power ratio at the primary receiver. Using integration by parts, \eqref{fd1} can be further simplified as
\begin{align}
\label{fd2}
\int^{\gamma_0}_{0}F(x)dx=\alpha
\end{align}
where $F(x)$ denotes the CDF of $g_0/g_1$. Hence, using \eqref{fd2} we have calculated $\gamma_0$ numerically. Note that the calculations in \eqref{first}-\eqref{fd2} are general and apply to both the equal power ($c=1$) and the unequal power ($c \ne 1$) case. In Section~III the required PDFs and CDFs were derived for the equal power case. When $c \ne 1$, it is a simple process to repeat the steps in  \eqref{first}-\eqref{fd2} and to show that using $c\alpha$ instead of $\alpha$ in \eqref{fd2} with the equal power results from Section~III gives the correct results. Hence, we only require the PDF and CDF of $g_1/g_0,g_0/g_1$ for the equal power case. To obtain results for the unequal power case, we simply use $c\alpha$ rather than $\alpha$ in \eqref{fd2}. This is equivalent to replacing $N_0$ by $N_0/c$, which makes intuitive sense since a power ratio of $c$ implies that the secondary receiver receives a signal $c$ times stronger than the primary. Hence, relative to the equal power case the SNR is $c$ times bigger and the equivalent noise level is $N_0/c$.

Figs. \ref{fig_2} and \ref{fig_3} show the secondary capacity versus $\alpha$ and under an average received interference power constraint. In all plots, AWGN refers to the scenario where $g_0$ and $g_1$ are equal to unity all the time \cite{ghasemi_b}. We make the following noteworthy observations:

\begin{figure}[t]
\centering
\includegraphics[width=0.95\columnwidth]{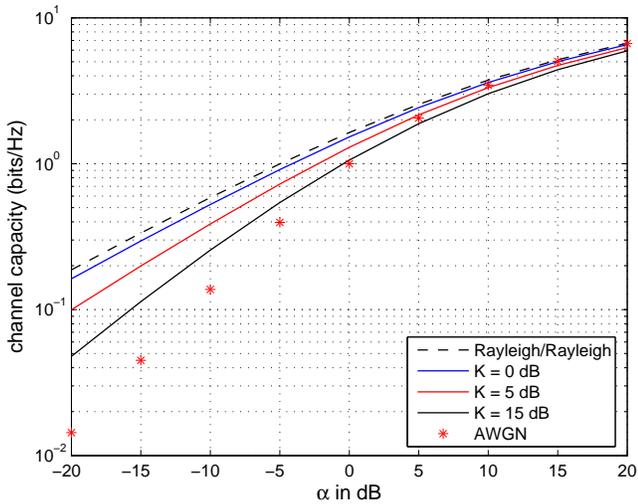}
\caption{Capacity under an average received-power constraint against $\alpha$ in Rayleigh/Rician fading. $c=0$ dB.}
\label{fig_2}
\end{figure}

\begin{figure}[t]
\centering
\includegraphics[width=0.95\columnwidth]{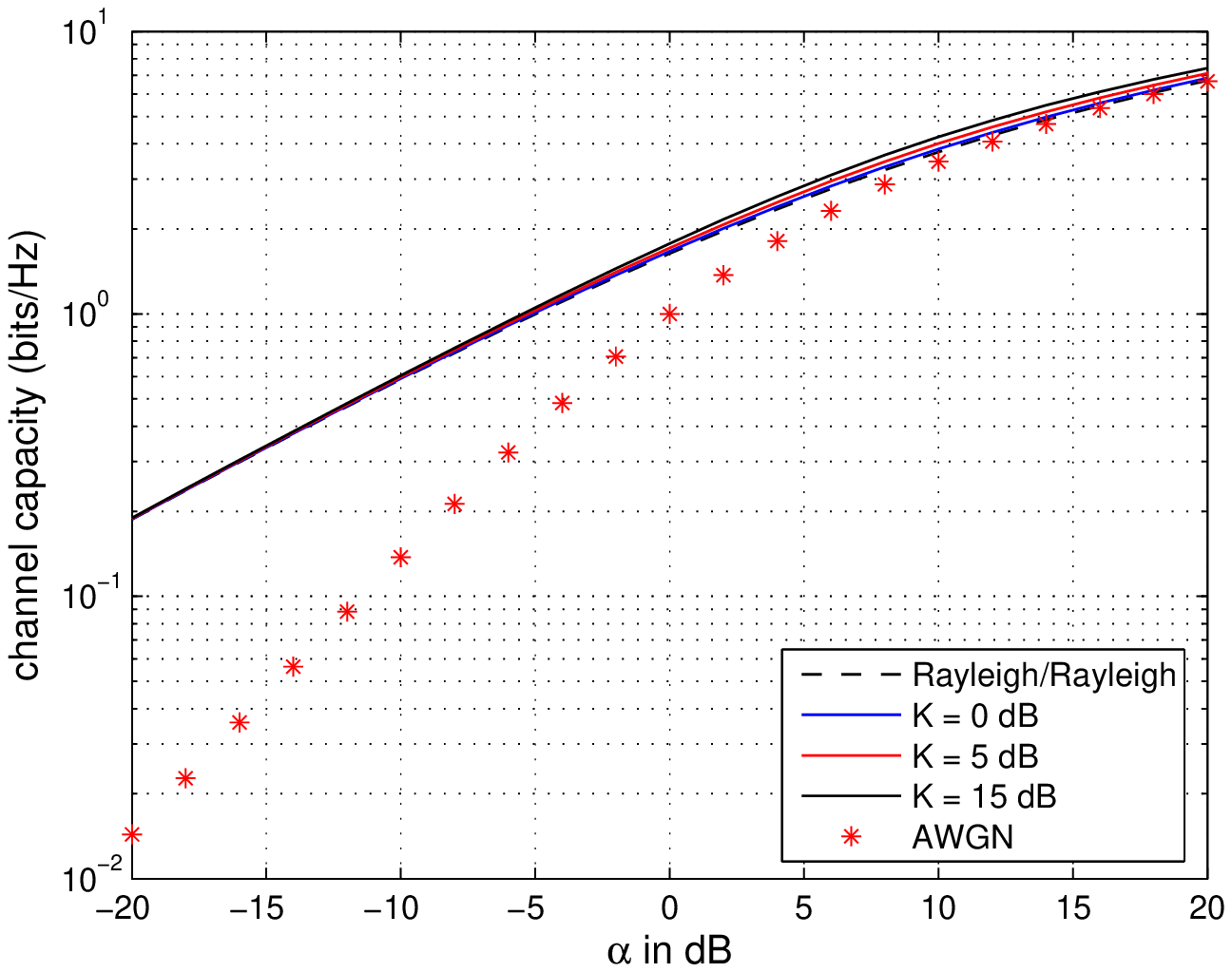}
\caption{Capacity under an average received-power constraint against $\alpha$ in Rician/Rayleigh fading. $c=0$ dB.}
\label{fig_3}
\end{figure}

\begin{figure}[t]
\centering
\includegraphics[width=0.95\columnwidth]{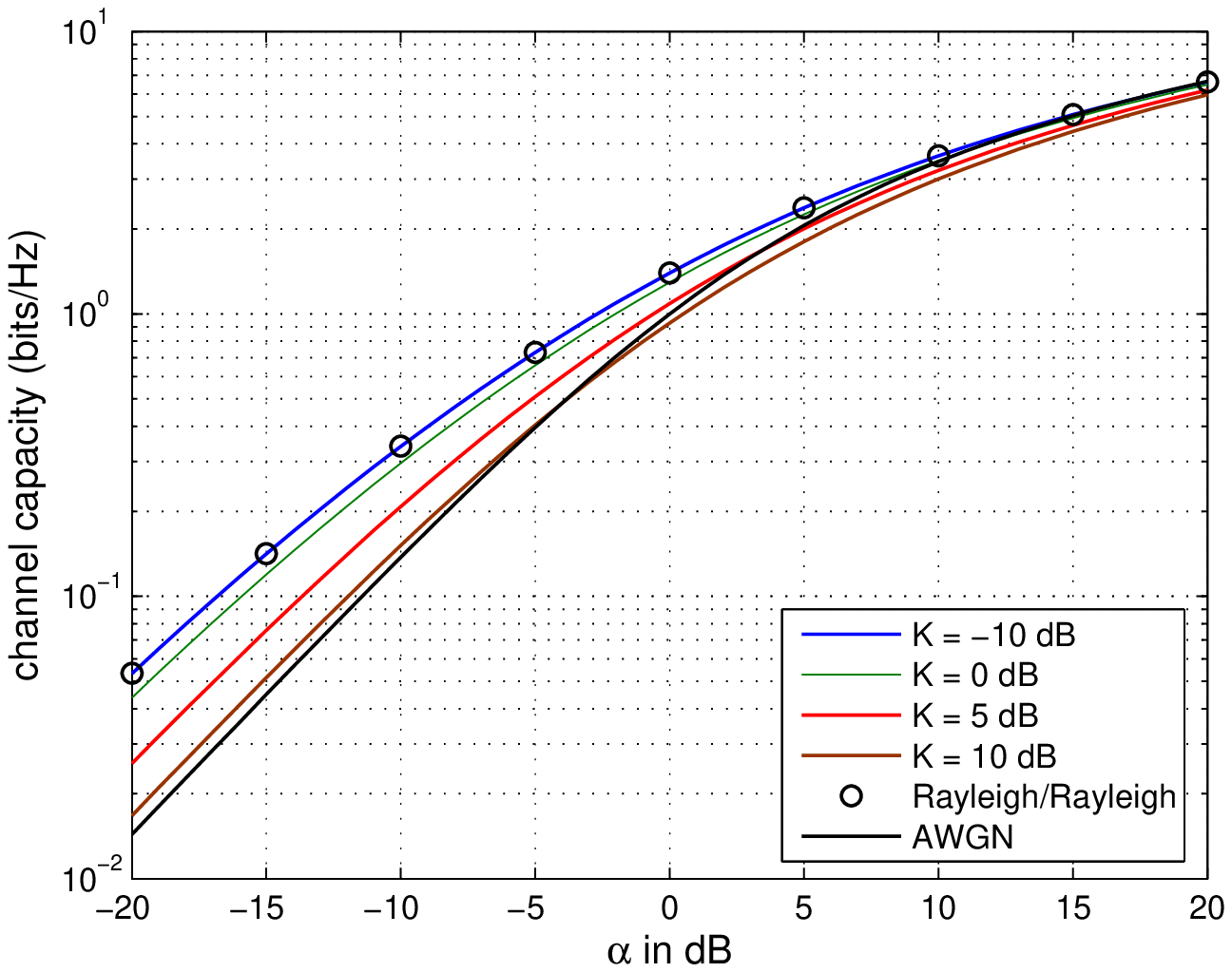}
\caption{Capacity under a peak received-power constraint against $\alpha$ in Rayleigh/Rician fading. $c=0$ dB.}
\label{fig_4}
\end{figure}

\begin{figure}[t]
\centering
\includegraphics[width=0.95\columnwidth]{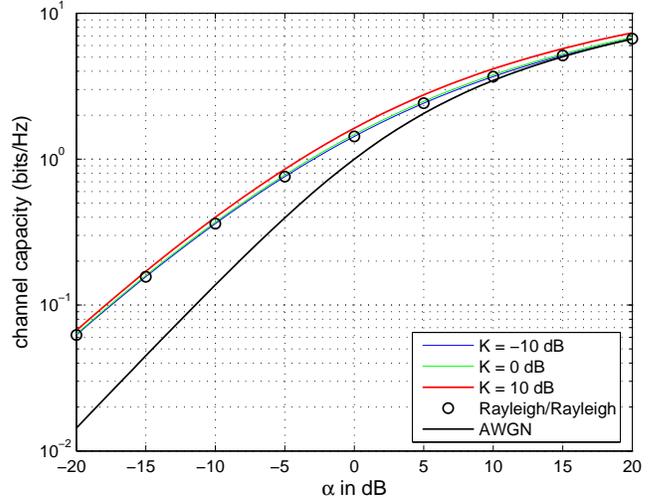}
\caption{Capacity under a peak received-power constraint against $\alpha$ in Rician/Rayleigh fading. $c=0$ dB.}
\label{fig_5}
\end{figure}

\begin{enumerate}

\item The secondary capacity increases if the primary receiver can tolerate more interference. This is because the secondary transmitter is able to transmit with higher power (probabilistically).

\item The case of interest in engineering practice is for a low value of $\alpha$, i.e., when the acceptable CR interference is correspondingly low. Here we see that the capacity can be sensitive to the type of fading and indeed the symmetric fading, i.e, the Rayleigh/Rayleigh case significantly overestimates the capacity compared to
the Rayleigh/Rician case in the low $\alpha$ regime. This observation is central to our contribution in
this paper. The difference in capacity reduces to almost zero when the acceptable interference at
the primary is large.

\item
The capacity of Rician/Rayleigh fading (cf. Fig. \ref{fig_3}) is not so sensitive to the $K$-factor $(0-15)$ dB. For a given $\alpha$, the Rayleigh fading on the primary link determines the transmit power of the secondary user. Once this is determined, the resulting secondary user capacity is less sensitive to the $K$-factor within
the considered range of $0-15$ dB. This is in contrast with Rayleigh/Rician fading, (cf. Fig. \ref{fig_4}), where we see that the $K$-factor induces an appreciable capacity difference especially in the low $\alpha$ regime. As $K$-factor decreases and in the low $\alpha$ regime, more opportunities for the secondary user to
transmit with relatively high power are created. However, for large $\alpha$, the impact of changing $K$-factor on the secondary user transmit power is reduced.

\item Under fading, the secondary capacity is higher than the AWGN case. This observation is consistent with the findings of \cite{ghasemi_b}. In a fading environment, the secondary user can transmit with high power, when its signal received by the primary user is subject to deep fades. 

\end{enumerate}
\subsection{Capacity Under a Peak Received-Power Constraint}
As discussed in \cite{ghasemi_b}, although \emph{average} received-power is reasonable for delay insensitive applications, in other cases it is desirable to impose a peak received-power constraint. Under the peak received-power constraint \cite{ghasemi_b}
\begin{equation}
g_{0}P(g_0,g_1)\leq Q
\end{equation}
and the channel capacity was given in \cite{ghasemi_b} as
\begin{equation}
\label{peakeq}
C=B\int_{0}^{\infty}\log_2\left(1+\alpha x\right)p_{\frac{g_1}{g_0}}(x)dx
\end{equation}
Therefore, under Rayleigh/Rician fading the channel capacity is obtained by substituting the PDF in \eqref{rayrician} into \eqref{peakeq}. This gives
\begin{align}
C = (K+1)B\int^{\infty}_{0} & \log_2\left(1+\alpha x\right) \frac{x+(K+1)^2}{(x+K+1)^3}\\\nonumber &
\times e^{-K+\frac{K^2+K}{x+K+1}} dx
\end{align}
\begin{figure}[t]
\centering
\includegraphics[width=0.935\columnwidth]{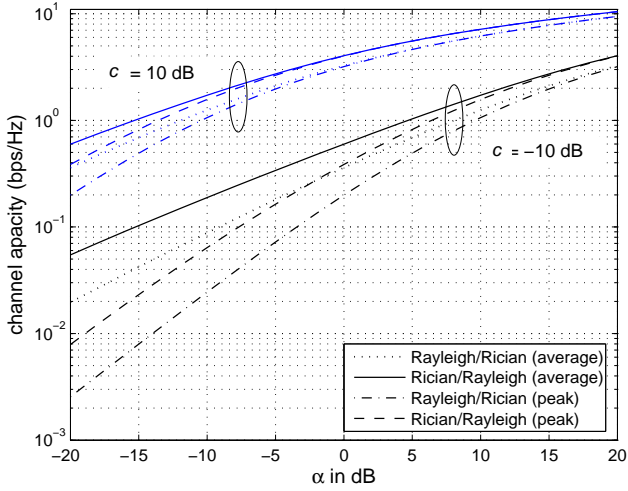}
\caption{Capacity under average and peak received-power constraints against $\alpha$ in Rician/Rayleigh and Rayleigh/Rician fading for two different values of $c$. $K=6$ dB.}
\label{fig_6}
\vspace{-3mm}
\end{figure}
\begin{figure*}[!t]
\normalsize
\setcounter{MYtempeqncnt}{\value{equation}}
\setcounter{equation}{22}
\begin{align}
\label{uy1}
C& =K(1+K)B \int^{\infty}_{0} \log_2\left(1+\alpha x\right) \frac{e^{-\frac{K}{x+Kx+1}}}{(x+Kx+1)^2}dx
 +(1+K)^{2}B\int^{\infty}_{0}\log_2\left(1+\alpha x\right)\frac{(1-K+x)}{(x+Kx+1)^3}e^{-K+\frac{Kx+K^2x}{x+Kx+1}}dx
\end{align}
\setcounter{equation}{\value{MYtempeqncnt}}
\hrulefill
\end{figure*}
\noindent Similarly, under Rician/Rayleigh fading the channel capacity is given by \eqref{uy1} on the next page. The case where the shadowing on the two links is different can be derived using the same arguments as above. Hence, numerical results are obtained assuming $g_1$ and $g_0$ have equal power but $\alpha$ is replaced by $c\alpha$.

Figs. \ref{fig_4} and \ref{fig_5} show the secondary capacity versus $\alpha$ and under a peak received interference power constraint. We make the following noteworthy observations:
\begin{enumerate}

\item  Like the average interference power case, the capacities increase if the primary can tolerate more interference.

\item The secondary capacity is sensitive to the type of fading on the two links and depending on the fading type on either link one could overestimate the capacity especially for low values of $\alpha$ and Rayleigh/Rician fading.

\item
From \cite[Fig. 4]{ghasemi_b} in symmetric fading conditions, the capacity under a peak received power constraint is always higher than the AWGN case. However in Rayleigh/Rician fading, we see that the capacity is higher/lower than the AWGN case depending on the $\alpha$.
\end{enumerate}

\begin{figure}[t]
\centering
\includegraphics[width=0.93\columnwidth]{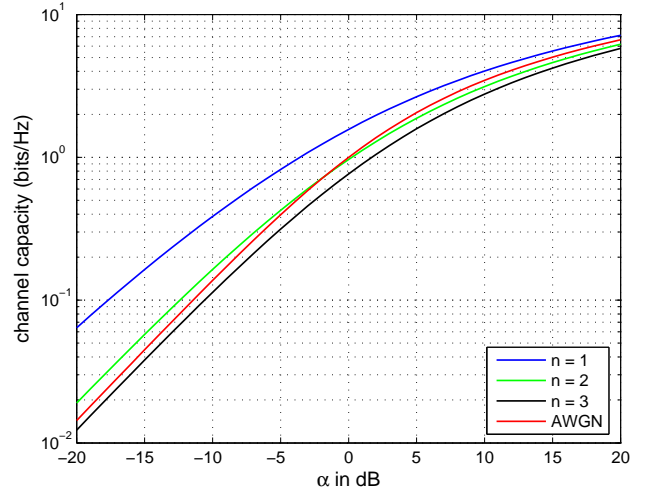}
\caption{Capacity under a peak received-power constraint and Rician/Rayleigh fading for different numbers of primary receivers. $K=6$ dB.}
\label{fig_7}
\end{figure}
\begin{figure}[t]
\centering
\includegraphics[width=0.93\columnwidth]{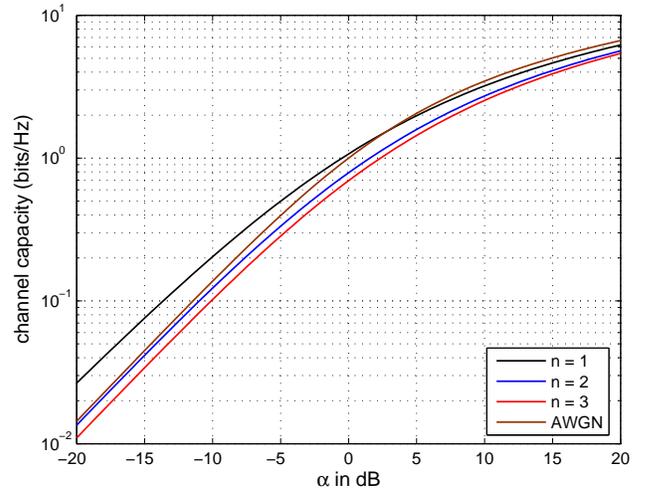}
\caption{Capacity under a peak received-power constraint and Rayleigh/Rician fading for different numbers of primary receivers. $K=6$ dB.}
\label{fig_8}
\vspace{-3mm}
\end{figure}

Fig. \ref{fig_6} shows the impact of signal power differences on CR capacity. Such differences usually arise from shadowing and  path length differences. We assume  two values for the power ratio between the links, $c=10$ dB and $c=-10$ dB. The effect on CR capacity is a simple scaling by the $c$ parameter. Hence, we have a simple and efficient approach to investigating such asymmetric links.
\begin{figure*}[!t]
\normalsize
\setcounter{MYtempeqncnt}{\value{equation}}
\setcounter{equation}{25}
\begin{align}
\label{uys1}
C=nB\sum^{n-1}_{k=0}(-1)^{k}\binom{n-1}{k}\int^{\infty}_{0}\log_{2}\left(1+\alpha x \right) \frac{1}{(1+k+(1+K)x)^2} e^{-\frac{(1+k)K}{1+k+(1+K)x}}\left(1+K+\frac{K(K+1)^2x}{1+k+(K+1)x}\right)dx
\end{align}
\setcounter{equation}{\value{MYtempeqncnt}}
\hrulefill
\end{figure*}
\addtocounter{equation}{1}
\section{Effect of Multiple Primary Users}
When $n>1$ primary users are present, the transmit/receive powers of the secondary user would be subject to additional constraints. This leads to a capacity reduction \cite{ghasemi_b}. Let $g_{0i}$ denote the channel gain of the secondary transmitter to the $i$-th primary receiver. In this case, the peak received-power constraint is reformulated by the following constraint
\begin{align}
P(g_{01},g_{02},\ldots,g_{0n},g_{1}) \leq \min_{i} \frac{Q}{g_{0i}},\ \ i=1,\ldots,n
\end{align}
The channel capacity is given by
\begin{align}
\label{capz}
C=B\int_{0}^{\infty}\log_{2}\left(1+\alpha z\right)p_{Z}(z)dz
\end{align}
where $Z=g_1/\max_{i} g_{0i}$. In Appendix A we have derived the PDF for $Z$ when $\sqrt{g_{0i}}$, $i,=1,\dots,n$ are independent and identically distributed (i.i.d) Rayleigh RVs and $\sqrt{g_1}$ is a Rician RV. Substituting this PDF into \eqref{capz} results in the capacity given by \eqref{uys1}. Such a result can be extended to the unequal power case by considering the maximum of independent Rayleigh variables with differing means. This is possible using standard order statistic results, but is beyond the scope of the paper.

Unfortunately, the PDF for the case when $\sqrt{g_{0i}}$, $i=1,\dots,n$ are i.i.d Rician RVs and $\sqrt{g_1}$ is a Rayleigh RV could not be found in closed-form. Instead we have resorted to \emph{time consuming} Monte-Carlo simulations to obtain the capacity. The average received-power case appears to be rather complex and is not considered here. Figs. \ref{fig_7} and \ref{fig_8} illustrate the CR capacity for $n=1,2,3$ and Rayleigh/Rician and Rician/Rayleigh fading respectively. In all cases, the capacity reduces compared to the AWGN case as $n$ gets larger.
\section{Conclusions}
In this paper we have investigated the impact of asymmetric fading on the secondary user
capacity under average and peak interference power constraints. Compared to symmetric fading conditions assumed in the previous literature, our analysis have added several new insights, especially for a low value of $\alpha$, i.e., the regime that most CRs would expect to operate in practice. The results show that under Rayleigh/Rician fading and low $\alpha$, the capacity is significantly lower than that in a symmetric Rayleigh/Rayleigh fading scenario, and as $\alpha$ increases, the impact of $K$-factor on the capacity is reduced. Under Rician/Rayleigh fading, the capacity results change only slightly with different $K$-factors within considered range of $0-15$ dB. The capacity results were also extended to include the effects of different power gain and multiple primary users. 
\appendices
\addtocounter{equation}{1}
\section{Derivations with multiple primary users}
Let $\sqrt{g_{0i}}$, for $i=1,\ldots, n$, be i.i.d Rayleigh RVs and let $\sqrt{g_1}$, which is independent of all $g_{0i}$, have a Rician distribution. Define $g_0=\max_{i} g_{0i}$ for $i=1,\ldots,n$ and $U=g_1/g_0$. Then the CDF of $U$ is given by
\begin{align}
\label{s1}
P(U<u)=\int^{\infty}_{0}P\left(g_1<g_0u|g_0\right)p_{g_0}(g_0)dg_0
\end{align}
The PDF of $g_0$, $p_{0}(g_0)$ is given by \cite[eq. 9.326]{simon} as
\begin{align}
\label{s2}
p_{g_0}(g_0)=n\sum^{n-1}_{k=0}(-1)^{k}\binom{n-1}{k}e^{-(1+k)g_{0}}
\end{align}
Substituting \eqref{s2} into \eqref{s1} we obtain
\begin{align}
\label{tr1}
P(U<u)& =1-n\sum^{n-1}_{k=0}(-1)^{k}\binom{n-1}{k}\\\nonumber
& \times \int^{\infty}_{0} Q_1\left(\sqrt{2K},\sqrt{2(1+K)ut}\right)e^{-(1+k)t}dt
\end{align}
After solving the integral in \eqref{tr1}, we express $P(U<u)$ in closed-form as
\begin{align}
F_{U}(u)&=1-n\sum^{n-1}_{k=0}\frac{(-1)^{k}}{1+k}\binom{n-1}{k}\\\nonumber
& \times \left(1-\frac{(1+K)u}{1+k+(1+K)u}e^{-\frac{(1+k)K}{1+k+(1+K)u}}\right)
\end{align}
Finally, differentiation of $P(U<u)$ with respect to $u$, yields the PDF of $U$. Therefore, the PDF of $U$ is given by
\begin{align}
p_{U}(u)&=n\sum^{n-1}_{k=0}(-1)^{k}\binom{n-1}{k}\frac{1}{(1+k+(1+K)u)^2} \\\nonumber & \times  e^{-\frac{(1+k)K}{1+k+(1+K)u}}\left(1+K+\frac{K(K+1)^2u}{1+k+(K+1)u}\right)
\end{align}
\section*{Acknowledgement}
This research is supported under the Australian Research Council's Discovery funding scheme (DP0774689).

\begin{thebibliography}{20}
\bibliographystyle{IEEEtran}
\bibitem{fcc}
Federal Communications Commission (FCC), ``Facilitating opportunities for flexible, efficient, and reliable spectrum use employing cognitive radio technologies,'' ET Docket No. 03-108, Mar. 2005.

\bibitem{ofcom}
Cognitive Radio Technology - A Study for Ofcom, [online] Available: \url{http://www.ofcom.org.uk/research/technology/research/emer_tech/cograd/cograd_main.pdf}
%
%
\bibitem{mitola}
J. Mitola III, \emph{Cognitive radio: An integrated agent architecture for software defined radio}, Ph.D Thesis, KTH Royal Institute of Technology, Sweden, May 2000.

\bibitem{haykin}
S. Haykin, ``Cognitive radio: Brain-empowered wireless communications,'' \emph{IEEE J. Select. Areas Commun.}, vol. 23, pp. 201-220, Feb. 2005.

\bibitem{jondral}
F. K. Jondral, ``Cognitive radio: A communications engineering view,'' \emph{IEEE Wireless Commun. Mag.}, vol. 14, pp. 28-33, Aug. 2007.

\bibitem{akyildiz}
I. F. Akyildiz, W.-Y. Lee, M. C. Vuran and S. Mohanty, ``Next generation/dynamic spectrum access/cognitive radio wireless networks: A survey,'' \emph{Computer Networks}, vol. 50, pp. 2127-2159, 2006.

\bibitem{kolodzy}
P. J. Kolodzy, ``Interference temperature: A metric for dynamic spectrum utilization,'' \emph{International Journal on Network Management}, vol. 16, pp. 103-113, 2006.


\bibitem{cabric}
D. Cabric, S. M. Mishra and R. W. Broderson, ``Implementation issues in spectrum sensing for cognitive radios,'' in \emph{Proc. Asilomar Conf. Signals, Systems and Computers}, pp. 772-776.

\bibitem{gastpar_1}
M. Gastpar, ``On capacity under received-signal constraints,'' in \emph{Proc. $42$nd Annual Allerton Conf. Communication, Control and Computing}, Monticello, IL, Sept.-Oct. 2004.

\bibitem{ghasemi_b}
A. Ghasemi and E. S. Sousa, ``Fundamental limits of spectrum-sharing in fading environments,'' \emph{IEEE Trans. Wireless Commun.}, vol. 6, pp. 649-658, Feb. 2007.



\bibitem{jafar}
S. A. Jafar and S. Srinivasa, ``Capacity limits of cognitive radio with distributed and dynamic spectral activity,'' \emph{IEEE J. Select. Areas Commun.}, vol. 25, pp. 529-537, Apr. 2007.




%
%
\bibitem{parsons}
J. D. Parsons, \emph{The Mobile Radio Propagation Channel}. New York: NY, Wiley, 1992.

\bibitem{prud}
A. P. Prudnikov, Y. A. Brychkov and O. I. Marichev, \emph{Integrals and Series}, vol. 2, Gordon and Breach Science Publishers, 1986.

\bibitem{nuttall}
A. H. Nuttall, ``Some integrals involving the $Q$-function,'' Naval Underwater Systems Center (NUSC), Technical Report 4297, April 1972.

\bibitem{nuttall_2}
A. H. Nuttall, ``Some integrals involving the $Q_M$-function,'' Naval Underwater Systems Center (NUSC), Technical Report 4755, May 1972.


\bibitem{simon}
M. K. Simon and M.-S. Alouini, \emph{Digital Communication over Fading Channels}, 1st ed. New York: Wiley, 2001.

%
%

\end{thebibliography}
\end{document}